\documentclass[apjl]{emulateapj} 	
\usepackage{apjfonts}			

\usepackage{natbib}
\usepackage{graphicx}
\bibpunct{(}{)}{;}{a}{}{,}

\shorttitle{
	Bright points in the inter-network quiet Sun
	}
\shortauthors{S\'anchez Almeida et al.}

\begin{document}
   \title{
	Bright points in the inter-network quiet Sun
	}

  \author{J. S\'anchez Almeida\altaffilmark{1},
	I. M\'arquez\altaffilmark{1,2},
	J. A. Bonet\altaffilmark{1},
        I. Dom\'\i nguez Cerde\~na\altaffilmark{3},
        {\sc and}
        R. Muller\altaffilmark{4}}
  \altaffiltext{1}{Instituto de Astrof\'\i sica de Canarias,
           E-38205 La Laguna, Tenerife, Spain;
  	jos@iac.es, 
	imr@iac.es, 
	jab@iac.es.
	   }
  \altaffiltext{2}{Departamento de An\'alisis Matem\'atico,
	Universidad de La Laguna.
	}
  \altaffiltext{3}{Universit\"ats-Sternwarte,
          Geismarlandstra\ss e 11, D-37083 
	  G\"ottingen, Germany;
	ita@uni-sw.gwdg.de. 
	}
  \altaffiltext{4}{Observatoire Midi-Pyrenees,
	14 Avenue Edouard Belin, 31400 
	Toulouse, France;
	muller@bagn.obs-mip.fr.
	}

\begin{abstract}
High resolution G-band images of the interior of a 
supergranulation
cell show ubiquitous Bright Points (some 0.3 BPs per Mm$^2$).
They  are located in intergranular lanes
and often form chains of elongated blobs 
whose smallest dimension  is at the resolution limit 
(135 km on the Sun). Most of them live for a few minutes,
having peak intensities from 0.8 to 1.8 times 
the mean photospheric intensity. 
These BPs are probably tracing intense
magnetic concentrations, whose existence has been 
inferred in spectro-polarimetric
measurements.
Our finding provides a new convenient tool for the study 
of the inter-network magnetism, so far restricted to the 
interpretation weak polarimetric signals.
%
\end{abstract}
\keywords{
          Sun: magnetic fields --
          Sun: photosphere}

%
%

\section{Introduction}\label{intro}

The presence of
Bright Points (BPs) in the solar photosphere 
was first reported by \citet{dun73} and \citet{meh74}. 
They were identified with small magnetic
concentrations, which  are expected to be bright \citep{spr77}\footnote{
The highly magnetized plasma of a magnetic 
concentration
is more transparent than the mean photosphere, allowing
photons to escape from deep (and usually) hot sub-photospheric
layers. 
}. 
BPs appear in all magnetic structures:
around sunspots, 
in plage regions, and in the network 
\citep[e.g.][]{mul94}.
This Letter shows how they are also very common 
inside supergranulation network cells.

The new result refers to the magnetism of the so-called Inter-Network (IN), 
and may be consequential.
The seemingly inactive IN provides
a large fraction of the (unsigned) magnetic flux and energy 
existing on the solar surface at any given time 
\citep[e.g.,][]{ste82,san02b,san04,tru04}. 
Therefore, it may be an important element to
understand the global magnetic properties of the Sun
(e.g., the structure of the quiet corona, \citealt{sch03b};
the chromospheric heating, \citealt{goo04}; the sources
of the solar wind, \citealt{woo97b,hu03};
or the emergence of magnetic
flux, \citealt{dep02}).
We still have a very primitive knowledge of the 
properties of the IN fields, 
which mostly comes from the model-dependent analysis of 
weak spectro-polarimetric signals 
\citep[for a recent review, see ][]{san04}.
The finding of G-band BPs provides a new simple
and direct mean to study the IN magnetism 
using unpolarized light.
In addition, the presence of BPs strongly suggests
that part of the IN fields have
kG field strengths. 
Spectro-polarimetric observations indicate IN
field strengths going all the way from zero to kG 
\citep[see, e.g., the  introduction of ][]{san03c}. 
The presence of kG field strengths has been particularly 
controversial, 
and this work provides independent support.
%
%



\section{Observations and image restoration}\label{observations}
\begin{figure*}
\epsscale{1.}
\plotone{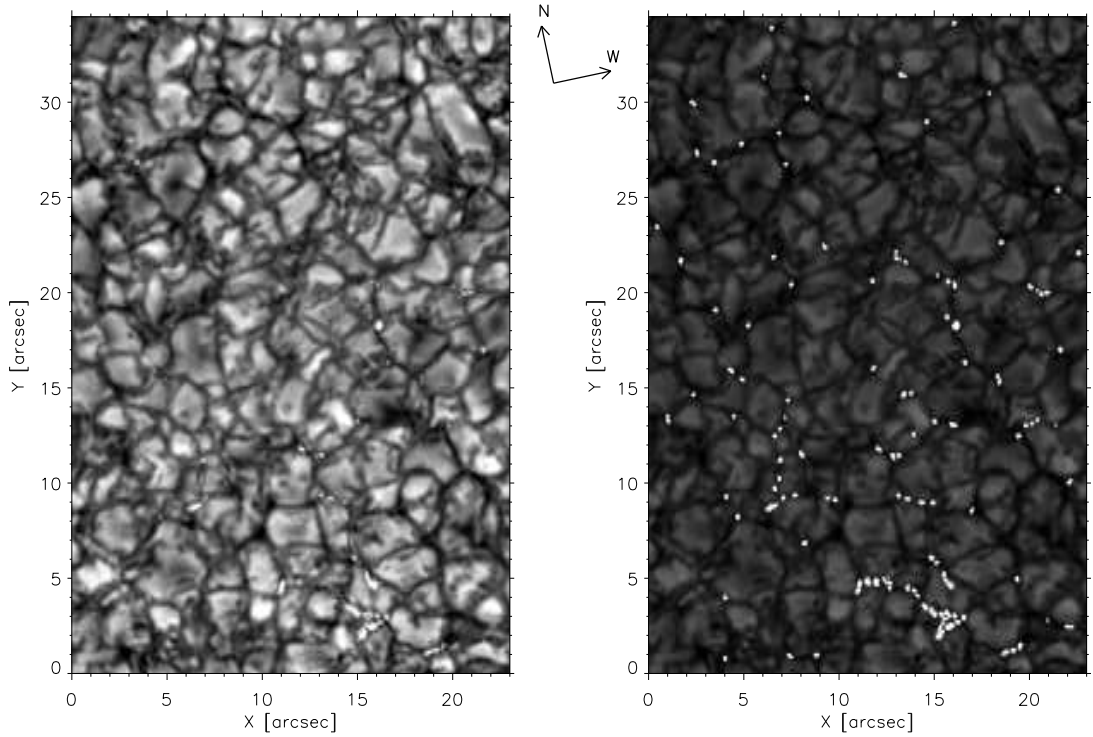}
\caption{Left: snapshot used for in-depth analysis. 
	Right: same snapshot with lower contrast and the 
	position of the selected BPs marked in white.
	The axes are in arcsec from the lower left corner.
	The arrows indicate the solar north and west directions during 
	observation. 
	}
\label{chickenpox}
\end{figure*}

A quiet Sun IN region was observed at the solar disk center
with the SST (Swedish Solar Telescope),
a new 1-m instrument equipped with adaptive
optics \citep{sch03c,sch03d}.
The Field-Of-View (FOV) was selected using Ca~{\sc ii}~H images to avoid
network magnetic concentrations.
We image through a 
10.8~\AA~wide filter centered in the G-band at 
4305.6~\AA . Although magnetic concentrations 
are bright at all wavelengths, they become
particularly conspicuous in this CH band
\citep{mul84,ber95}.
The optical setup included a Phase Diversity (PD) device for
post-facto restoration of the optical aberrations
induced by the terrestrial atmosphere and the telescope
\citep{gon79,pax92}.
The series of PD images  was taken 
on September 30, 2003, from 11:33~UT to 12:23~UT. We use a
10-bit Kodak Megaplus 1.6 camera (pixel size 0\farcs 041,
exposure time 8 msec).  Real-time frame selection provided us with the 
4 sharpest images every  20~sec, which are needed in our PD 
inversion code
(\citealt{bon03}, based on
\citealt{lof94}, and \citealt{pax96}).
%
%
First, each image of the series is divided in a mosaic of 96 
overlapping isoplanatic patches, which we restore independently.
Then the mosaic is assembled, and 
the 4 images of each selection interval are combined to form 
a restored snapshot. The process of splitting,
restoring and merging renders a 
FOV of 23\arcsec x 35\arcsec; see
Fig.~\ref{chickenpox}.
Excellent and rather bad moments of seeing alternated  
during the observing run. Therefore  
several sets of 4 images could 
not be successfully restored, leaving gaps in our series.
It finally has 100 restored snapshots during its 50~min duration.
The PD code includes a self-regulated high-spatial-frequency noise 
removal filter. 
Given the conditions of the data, the largest
cutoffs set by this filter are 0\farcs14, i.e., only slightly worst than the
diffraction limit (0\farcs 09). Note that the cutoff
differs for each PD restoration, and so, it varies  within
a single restored snapshot.
The cutoff approximately yields the Full Width Half Maximum
(FWHM) of the point spread function,
a rule of thumb used in \S~\ref{twogauss}.

\begin{figure}
\epsscale{1.0}
\plotone{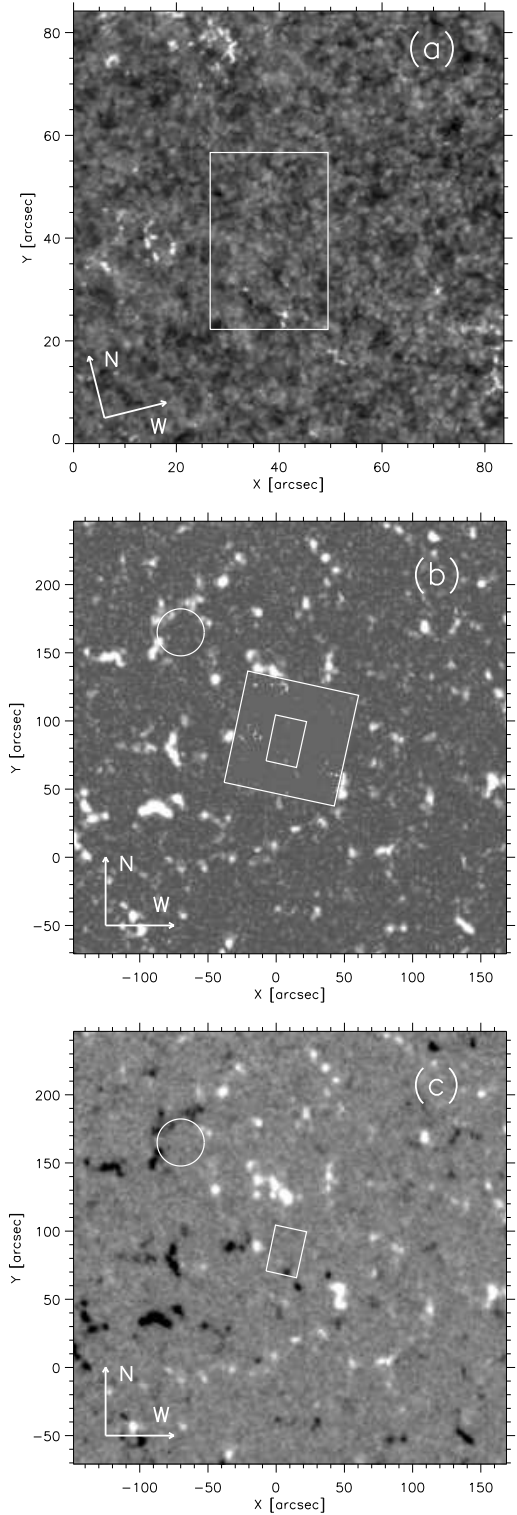}
\caption{(a) Ca~{\sc ii}~H image corresponding to the reference G-band snapshot.
	The position of the G-band FOV  within this larger Ca image is
	marked by the box. Axes are in arcsec from the lower left
	corner.
	(b) Absolute value of an MDI magnetogram taken right before
	the observation.
	The inset
	corresponds to the Ca image in (a), and it reveals
	MDI network patches continuing  within
	the SST Ca image. Axes are in arcsec
	from the solar disk center. The circle shows
	the typical size of a supergranulation cell.
	(c) Section of the MDI magnetograms scaled between $\pm 50$ G.
	The box corresponds to the G-band FOV. 
	%
	}
\label{ines}
\end{figure}

	Ca images allowed us to center
the G-band FOV within a supergranulation cell.
Figure \ref{ines}a shows one of these Ca images, which includes
a box with the G-band FOV outline. Note how the brightest Ca
patches remain outside the G-band contour. Only a small patch  
in the lower right corner lies within the bounds. 
In order to stress the IN character of our target, 
we pinpoint the observed FOV in a
full disk MDI magnetogram\footnote{http://soi.stanford.edu}
taken 18 minutes before the beginning of the 
series.
It is shown in Fig.~\ref{ines}b.
The figure includes an inset with the SST Ca image of the region.
%
Figure~\ref{ines}b was set up to emphasize the continuity
between the MDI magnetogram and the Ca images, and so,
it displays the unsigned polarimetric
signals.
Figure~\ref{ines}c shows the true MDI magnetogram saturated at
$\pm~50$~G to reveal the weakest network features. 
A small blob appears within the
G-band~FOV. The
rest remains field-free as far as the MDI magnetogram
is concerned. The MDI signals within the FOV
are $(-0.9\pm 7.1)$~G, with the 
standard deviation matching
the  MDI noise (e.g., 6.9~G for \citealt{hag01}).
%
%
The magnetic patch of the MDI magnetogram
is very conspicuous in the G-band; see 
the conglomerate of BPs at the lower right
part of Fig.~\ref{chickenpox}, coordinates [16\arcsec,3\arcsec]. 

\section{Analysis and Results}\label{all}

\subsection{Area coverage and number density}\label{byeye}
The sharpest snapshot of the time series was selected for in-depth 
study (Fig.~\ref{chickenpox}, left). Although this reference
snapshot was chosen by visual inspection, it has one of the 
largest 
contrasts of the full series (14.3\%).
Then, playing the series back and forth,
we single out all those BPs which (a) were in the reference 
snapshot, (b) persisted three or more snapshots, and
(c) stay in an intergranular lane. 
This subjective selection
of BPs was complemented with an automatic determination 
of their areas. We apply the segmentation algorithm  by
\citet[][ Chapter~8]{str94},
and then those patches overlaying a visually
selected BP were chosen as the area of the BP.
The result is shown in Fig.~\ref{chickenpox}, right.
We identify 126 individual 
points, some of which are part of long chains.
Given the area of the FOV, we detect
0.3~points~per~Mm${^{2}}$.
The density of BPs increases towards the lower left
part of the FOV (Fig.~\ref{chickenpox}), i.e., when approaching
the network patch (Fig.~\ref{ines}c).
The area associated 
with these BPs covers some 0.5\% of the FOV.
This area depends on the segmentation algorithm,
and a more meaningful
estimate is carried out in
\S~\ref{twogauss}.

\subsection{Size and Brightness}\label{twogauss}

\begin{figure}
\epsscale{1.}
\plotone{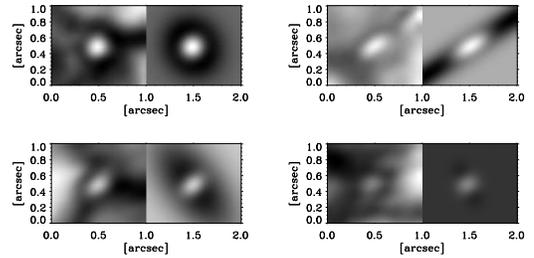}
\caption{Four representative pairs of 2-Gaussian
	fits (right) and the corresponding observed G-band BPs (left).
	}
\label{example}
\end{figure}

	The BPs stay in intergranular lanes, which are
narrow dark wells of the intensity distribution. \citet{tit96} show how
this fact biases any measurement of size unless the shape of
the background is considered.
They propose a 2-Gaussian fit to decontaminate from 
the influence of the intergranular background. 
We adopt this approach for
estimating the widths of our BPs. Those pixels
selected by the segmentation algorithm were
fitted using 
\begin{eqnarray}
&f=c+aG(x,y,x_0,y_0,\sigma_x,\sigma_y,\theta)
	-bG(x,y,x_0,y_0,\sigma^\prime_x,\sigma^\prime_y,\theta),\cr
	\label{eq1}
&G(x,y,x_0,y_0,\sigma_x,\sigma_y,\theta)=
\exp{-(r_x^2+r_y^2)},\cr
&r_x=[(x-x_0)\cos\theta+(y-y_0)\sin\theta]/\sigma_x,\cr
&r_y=[-(x-x_0)\sin\theta+(y-y_0)\cos\theta]/\sigma_y.
\end{eqnarray}
The 2-Gaussian function  $f$ depends on the spatial
coordinates $x$ and $y$, as well as on 10 free parameters:
$a, b,$ and $c$ for the amplitudes of two Gaussians and
a background, $x_0$ and $y_0$ for the core of the BP,
$\theta$ for a global orientation,
$\sigma_x$ and $\sigma_y$ for the widths of the 
BP and, finally, $\sigma^\prime_x$ and $\sigma^\prime_y$
for the widths of the 
intergranular lane. 
By means of a non-linear least squares algorithm, we
tried to reproduce to all BPs selected in Sec.\ref{byeye}.
Some 60\% of the trials converged.   
Figure \ref{example} shows four representative cases, which
indicate how the elliptical contours of 
the 2-Gaussian functions really reproduce
the observed BP shapes. 

Figure \ref{tripode}a contains the
histograms of the FWHM 
of the 2-dimensional Gaussians representing the BP 
(FWHM~=~$1.67\sigma_x$ or $1.67\sigma_y$). 
The minor axis FWHM peaks at some 135~km, 
and it seldom exceeds 200~km.
We believe that this 135~km represents the mean spatial
resolution of the snapshot. As we discuss in
\S~\ref{observations}, the resolution varies along the
FOV, with a lower limit of some 0\farcs 14 or 100~km.
The width 
of the minor axis histogram is mostly set by uncertainties
in the non-linear least squares fit
(some 30 km, as provided by the
fitting routine).

The interpretation of the areas derived in \S~\ref{byeye} is
ambiguous.
We prefer a straight definition
based on the 2-Gaussian fits, namely, the area within the elliptical 
contour embraced by the FWHM (i.e., $2.18\sigma_x\sigma_y$).
This definition yields areas 40\% larger than the 
segmentation algorithm. Then 
the 0.5\% coverage in \S~\ref{byeye} turns out to be   0.7\%.

Figure \ref{tripode}b shows a scatter plot of the
peak G-band intensity 
(i.e., $a-b+c$ in Eq. [\ref{eq1}])
versus minor axis FWHM.
There is no obvious correlation between brightness
and size, which can be interpreted
as the unresolved character of most of the features
\citep[e.g.,][]{ber95}. Figure \ref{tripode}b also shows many BPs with
an intensity smaller than that of the mean photosphere.
On top of this,
a few BPs reach 1.8 \citep[close but still below
the expected intensity for a fully resolved magnetic concentration; ]
[]{kis01,san01,stei01,sch03e}. 
In a different snapshot  of the series, 
the network clump peak intensity scores 2.3.

\subsection{Lifetimes}
\begin{figure}
\epsscale{1.0}
\plotone{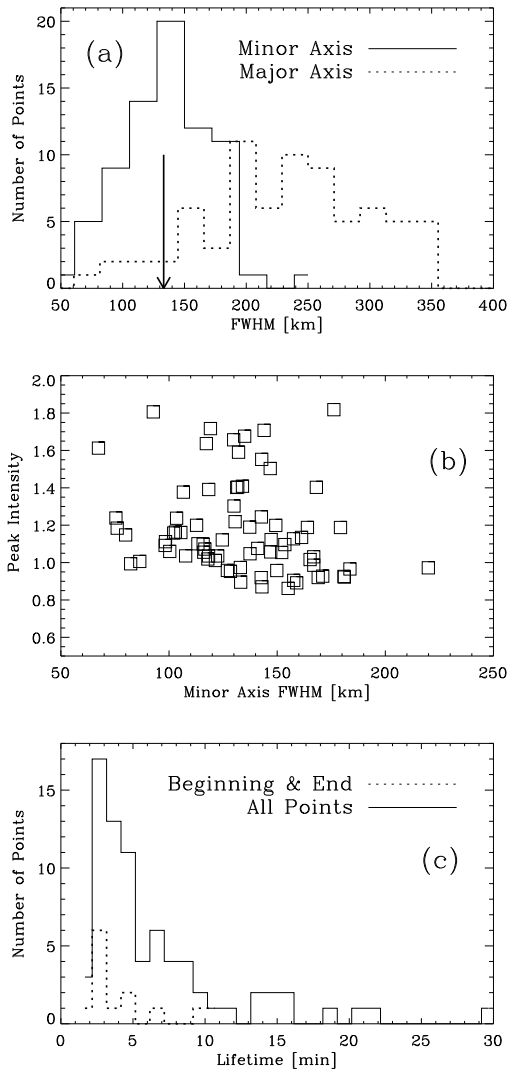}
\caption{(a) Histograms of the G-band BP FWHM obtained from the 
	2-Gaussian fits. The mean FWHM along the minor 
	axis is 135~km (see the arrow
	on the figure), which we interpret as the
	spatial resolution of the observation. 
	(b) Peak intensity versus FWHM. No obvious correlation
	exists. The intensities are refereed to the
	mean photosphere. (c) Histograms
	of G-band BP lifetimes. The dotted line is based on those
	BPs whose beginning and end we witnessed.
	}
\label{tripode}
\end{figure}

Lifetimes were estimated by visual inspection of the time series.
The BPs in the reference snapshot were visually tracked to 
find out when they appear and disappear. This time interval
defines the lifetime. A histogram of lifetimes is represented
in Fig.~\ref{tripode}c, the solid line. Most
lifetimes are shorter than 10~min.
In addition, some BPs live as much as we can measure (longer
than 30~min). Keep in mind that the lifetimes
are strongly biased.
First, BPs often appear or end 
within one of the gaps of the time series, which 
underestimate the true values. Second, the
detection criteria 
bias the estimates towards long-lasting BPs
(\S~\ref{byeye}).
Although these caveats should be kept in
mind, we believe that 
most BPs are really short-lived. Figure~\ref{tripode}c,
the dotted line, shows a histogram based on BPs whose
birth  and death we witnessed, and they have short
lifetimes.

\section{Comments and conclusions}\label{conclusions}

We have detected many Bright Points (BPs) in the interior
of a supergranulation cell (i.e., in the 
Inter-Network or IN). 
According to the current paradigm, these
bright points trace intense magnetic concentrations 
(\S~\ref{intro}). 
Our finding has two main consequences. First,
it provides a new convenient tool for the study 
of the IN magnetism, so far restricted to the interpretation
of low spatial resolution weak polarimetric signals.
Second, it supports a result based on 
Zeeman magnetometry indicating that part of the IN magnetic
fields are not weak but have kG magnetic field 
strengths 
\citep{san00,soc02,soc03}. The same
polarimetric measurements also point out how these
kG concentrations carry the body of
the IN magnetic flux and energy 
\citep{san03c,san04}. Consequently,
unpolarized imaging may allow us to
detect and study a significant fraction
of the IN magnetism.

The detected BPs cover 
some 0.7\% of the surface, which is not enough to account for 
the magnetic signals found by 
(\citealt{dom03a},
\citeyear{dom03b},
\citealt{san03d}). 
Several reasons may explain this difference.
We miss many BPs
since the detectability critically depends on the 
resolution \citep{tit96}, and our BPs remain unresolved
(\S~\ref{twogauss}). 
On the other hand, some strong fields may not be bright.
The actual brightness depends on subtleties of the magnetic
concentration and its non-magnetic environment,
and sometimes modeling shows faint kG  features
\citep{san01}.
Yet another possibility is the random
fluctuation of properties of different
cell interiors.
%


%

\acknowledgements
The work was funded by the Spanish 
project AYA2001-1649 and
the EC contract HPRN-CT-2002-00313.
Thanks are due to 
F.~Kneer for continuous support, and to
A.~Sainz~Dalda for help with MDI data
(courtesy of the SOHO/MDI consortium by ESA \& NASA). 
The SST is operated by the Institute for Solar Physics, Stockholm,
at the ORM of the IAC.
%
%
%

%

\end{document}